\documentclass[sigconf]{acmart}
\usepackage{subcaption} 
\AtBeginDocument{%
  }
 \usepackage{multirow}
\usepackage{algorithm}
\usepackage{algpseudocode}
\begin{document}

\title{Adaptive Command: Real-Time Policy Adjustment via Language Models in StarCraft II}

\author{
  {\bf Weiyu Ma$^{1,2}$},
  {\bf Dongyu Xu$^{1,2}$},
  {\bf Shu Lin$^{1}$},
  {\bf Haifeng Zhang$^{1,2,4}$}$^\dag$,
  {\bf Jun Wang$^{3}$}$^\dag$
  \vspace{0.05 cm}\\
  {\normalsize $^1$ Institute of Automation, Chinese Academy of Sciences, China}\\
  {\normalsize $^2$ School of Artificial Intelligence, University of Chinese Academy of Sciences, China}\\
  {\normalsize $^3$ AI Centre, Department of Computer Science, UCL}\\
  {\normalsize $^4$ Nanjing Artificial Intelligence Research of IA, China}\\
  {\footnotesize $^\dag$Correspondence to: $\langle$haifeng.zhang@ia.ac.cn$\rangle$, $\langle$jun.wang@cs.ucl.ac.uk$\rangle$}
}

\renewcommand{\shortauthors}{Ma et al.}


\begin{abstract}
We present Adaptive Command, a novel framework integrating large language models (LLMs) with behavior trees for real-time strategic decision-making in StarCraft II. Our system focuses on enhancing human-AI collaboration in complex, dynamic environments through natural language interactions. The framework comprises: (1) an LLM-based strategic advisor, (2) a behavior tree for action execution, and (3) a natural language interface with speech capabilities. User studies demonstrate significant improvements in player decision-making and strategic adaptability, particularly benefiting novice players and those with disabilities. This work contributes to the field of real-time human-AI collaborative decision-making, offering insights applicable beyond RTS games to various complex decision-making scenarios.
\end{abstract}



\maketitle

\section{Introduction}

\label{sec:introduction}

\begin{figure}[t]
  \centering
  \begin{subfigure}[b]{0.50\textwidth}
    \includegraphics[width=\textwidth]{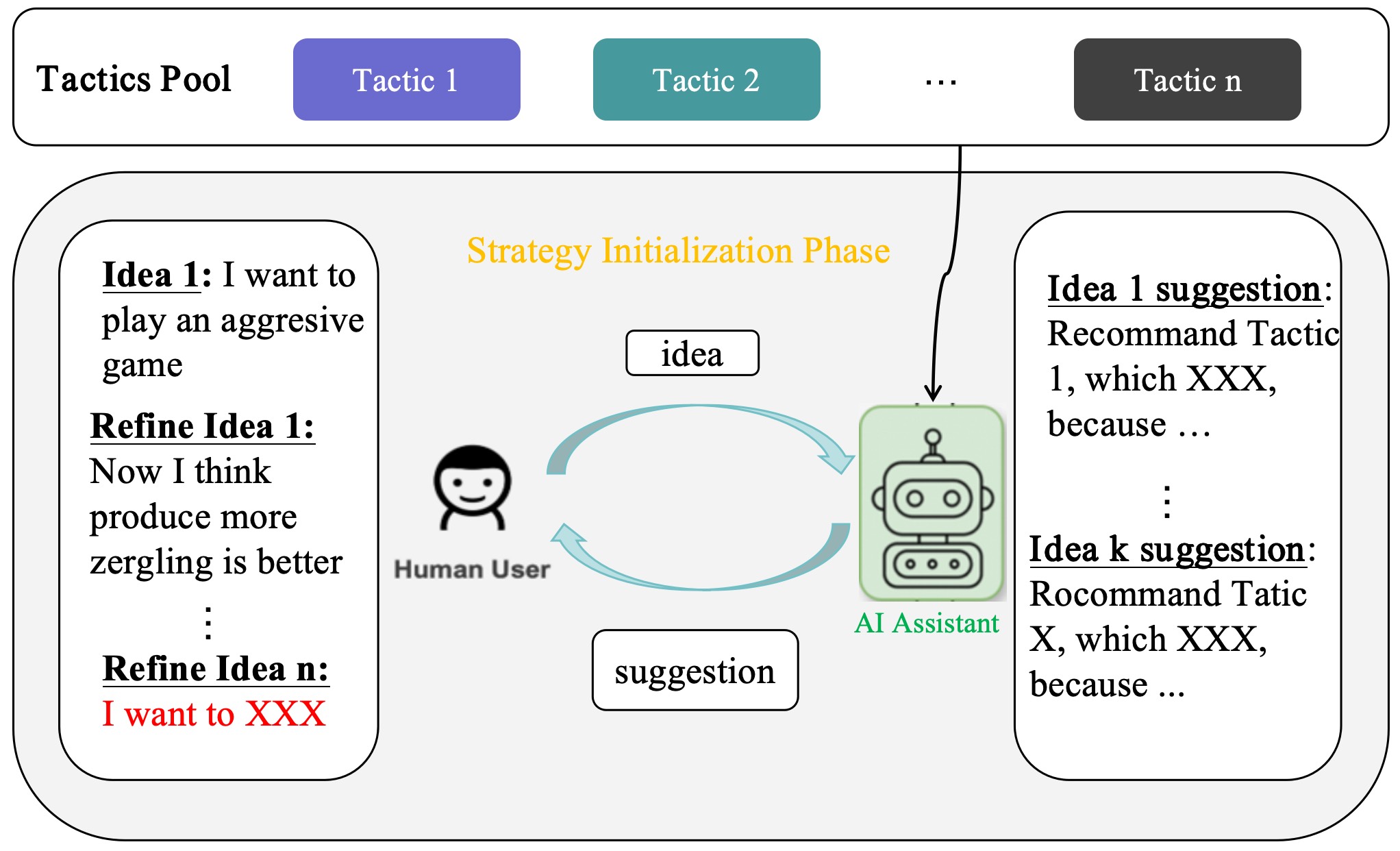}
    \caption{Strategy Initialization Phase}
    \label{fig:stage1}
    \Description{Diagram showing the Strategy Initialization Phase of Adaptive Command. It illustrates the initial interaction between the player, LLM, and the game environment to establish the starting strategy.}
  \end{subfigure}
  \hfill
  \begin{subfigure}[b]{0.50\textwidth}
    \includegraphics[width=\textwidth]{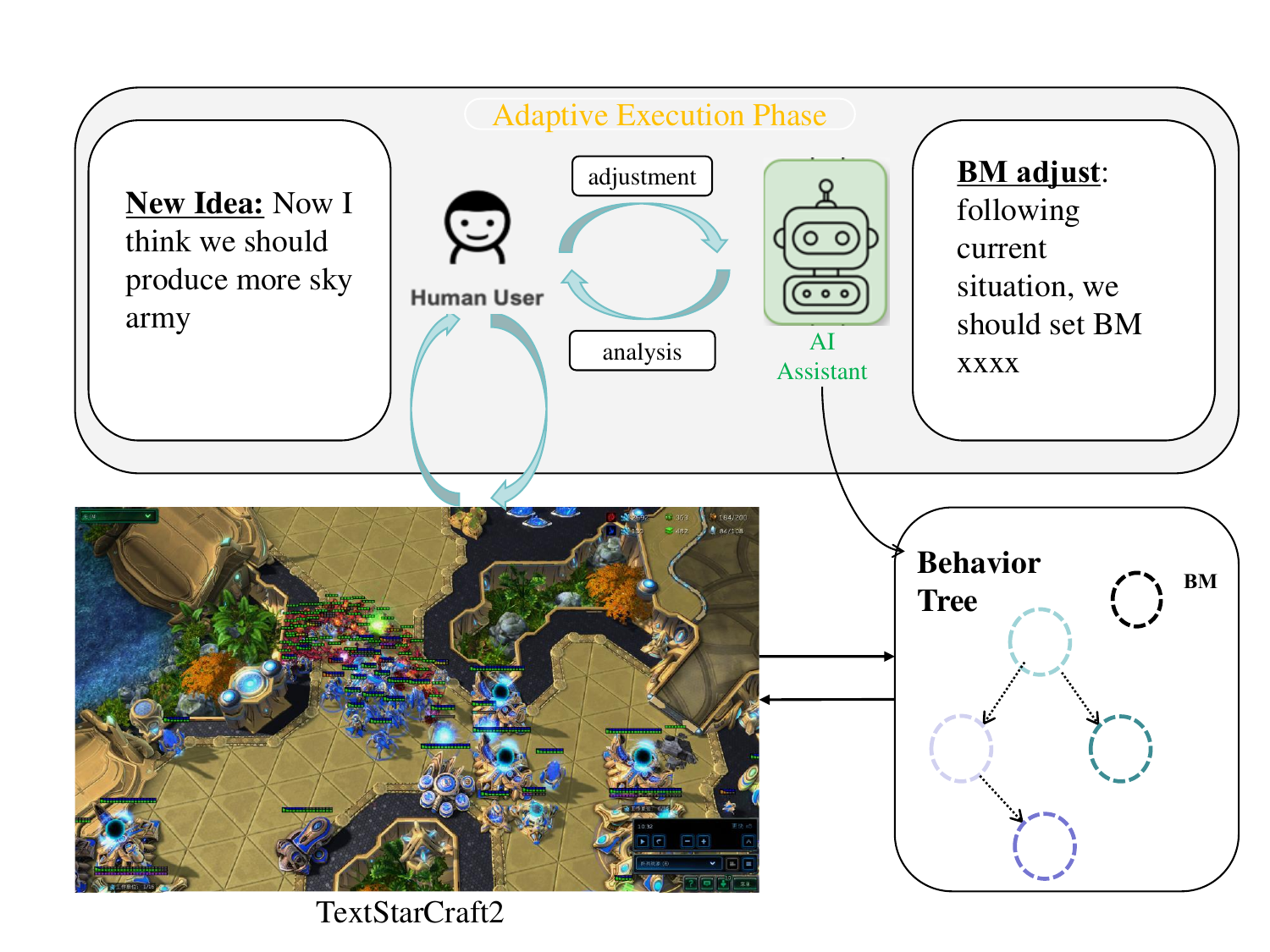}
    \caption{Adaptive Execution Phase}
    \label{fig:stage2}
    \Description{Diagram depicting the Adaptive Execution Phase of Adaptive Command. It shows the continuous loop of gameplay, player input, LLM-based strategy adjustment, and behavior tree execution.}
  \end{subfigure}
  \caption{Adaptive Command Framework: Two-stage process integrating LLMs with behavior trees for strategic decision-making in StarCraft II.}
  \label{fig:adaptive_command_framework}
  \Description{A two-part figure illustrating the Adaptive Command Framework. The left subfigure shows the Strategy Initialization Phase where the initial game strategy is formulated. The right subfigure depicts the Adaptive Execution Phase, showing how the strategy is continuously updated during gameplay based on player input and game state.}
\end{figure}

Real-time strategy (RTS) games, exemplified by StarCraft II (SC2), present formidable challenges in AI research, demanding rapid tactical decisions and adaptive long-term planning. SC2's complex gameplay, detailed in Section \ref{appendix:introduction of starcraft2}, encompasses resource management, base construction, and military command, serving as an ideal testbed for advanced AI systems. While previous efforts, such as DeepMind's AlphaStar \citep{alphastarnature}, have shown AI's potential in mastering such games, recent advancements have begun to explore the integration of Large Language Models (LLMs) in this domain.

The rapid advancement of LLMs in reasoning, planning, and decision-making has opened new possibilities for their application in complex gaming environments. A notable contribution in this area is the TextStarCraft II project \citep{ma2024largelanguagemodelsplay}, which developed a specialized environment for assessing LLMs in real-time strategic scenarios within SC2. Their work demonstrated that fine-tuned LLMs could perform on par with Gold-level players in real-time matches, showcasing the potential of LLMs in strategic gameplay.

Building upon these developments, we identify a gap in frameworks that leverage LLMs for enhancing human-AI collaboration in real-time strategy decision-making and long-term planning in environments like StarCraft II. While TextStarCraft II focused on evaluating LLMs' autonomous performance, our work aims to explore how LLMs can augment human gameplay through real-time interaction and collaboration.

To address this gap, we present Adaptive Command, a system based on TextStarCraft II that extends the capabilities of LLMs in StarCraft II beyond autonomous play to facilitate human-AI collaboration. Our approach introduces a language-conditioned and adjustable StarCraft II policy, leveraging the strengths of LLMs in reasoning and strategic planning, while integrating them with behavior trees for tactical execution.

Adaptive Command introduces several key innovations:

\begin{itemize}
    \item \textbf{Language-Conditioned Policy}: Integration of LLMs with behavior trees provides a flexible policy that adapts to player input and game state.
    \item \textbf{Real-time LLM-assisted Decision Making}: The system offers adaptive, context-aware strategic advice during gameplay, enhancing player decision-making capabilities.
    \item \textbf{Natural Language Interface}: Players interact with the AI assistant using natural language, including voice commands, facilitating intuitive strategic discussions.
    \item \textbf{Dynamic Strategy Adaptation}: Real-time policy adjustments allow players to modify strategies based on evolving game conditions and AI recommendations.
    \item \textbf{Enhanced Accessibility}: Incorporation of speech-to-text and text-to-speech capabilities aims to broaden the accessibility of complex RTS gameplay.
\end{itemize}

Our research investigates how this language-conditioned, human-AI collaborative approach can enhance player decision-making, strategic flexibility, and overall game performance. Through comprehensive user studies, we evaluate the system's effectiveness across various player skill levels and assess its potential in improving gaming experiences.

This work contributes to the broader field of human-AI collaboration in complex, real-time decision-making environments. While centered on StarCraft II, the insights and methodologies developed have potential applications beyond gaming, offering new perspectives on integrating AI assistance in dynamic, strategic scenarios across various domains. By focusing on the collaborative aspect between humans and AI, our work extends the capabilities of LLM-based systems in RTS games, potentially opening new avenues for AI-assisted gameplay and decision-making in complex environments.

\section{Related Work}
\label{sec:related_work}

\textbf{StarCraft II Full Game AI}: StarCraft AI research, initially focused on StarCraft I with developments like BiCNet \cite{peng2017multiagent} for multi-agent coordination, has significantly advanced in the StarCraft II era. The release of PySC2 \cite{pysc2} by DeepMind, coupled with Blizzard's game replays, propelled this research field. A key breakthrough was AlphaStar \cite{alphastarnature}, which achieved Grandmaster level and defeated top players, demonstrating the potential of RL in complex environments.

Subsequent research \citep{wang2021scc, christianos2021scaling, jiang2021multi, hu2024learning} expanded upon these foundations. Mini-AlphaStar\cite{liu2021introduction} simplified input variables without compromising learning effectiveness. TG\cite{liu2021efficient} and HierNet-SC2\cite{liu2022onefficient} explored efficient RL strategies, with the latter bypassing supervised pre-training. AlphaStar Unplugged\cite{starcraft2unplugged} represented a leap in offline RL using human replays. TStarBotsX\cite{Tstarbot-x} and SCC\cite{scc} furthered federated learning approaches, achieving notable success against master and grandmaster level players.

Recent advancements include DI-star\footnote{\url{https://github.com/opendilab/DI-star}}, which is accessible for home computer deployment, and ROA-Star\cite{ROA-Star}, enhancing AlphaStar's training framework with goal-conditioned exploiters and refined opponent modeling techniques. ROA-Star's practical tests against professional players have shown impressive results, marking significant progress in real-time strategy AI.

Grounding Natural Language Commands to StarCraft II\citep{groundingstar} made a pioneering contribution in this direction with their work on Grounding Natural Language Commands to StarCraft II Game States. 

Building upon this integration of language and game mechanics, TextStarCraft II \citep{ma2024largelanguagemodelsplay} further advanced the field by introducing a specialized environment for evaluating Large Language Models (LLMs) in real-time strategic scenarios within SC2. This work highlights the potential of language models in strategic gameplay and decision-making.

These advancements highlight the evolution of StarCraft II AI research, integrating natural language understanding with traditional approaches. This progression aims to create AI systems that interact with complex strategic environments in more human-like ways.

\textbf{LLM Agents and Benchmarks}: The surge of large language models (LLMs) in late 2022, exemplified by models such as , GPT-3.5,GPT4 \citep{gpt4_report}, and subsequent developments like LLaMA \citep{llama3} and Qwen \citep{bai2023qwen,yang2024qwen2}, has significantly propelled research into LLM agents. Projects like React , Chain of Thought \citep{react,chainofthought,chainofthoughtsc}, and AutoGPT laid the groundwork for more sophisticated implementations. In gaming environments, initiatives such as GITM \citep{ghostintheminecraft} and Voyager \citep{voyager} have demonstrated LLM agents' adaptability to various tasks and open-world scenarios. Environments like TextWorld \citep{cote2019textworld} and ALFWorld \citep{shridhar2020alfworld} have integrated text-based strategy and action execution. Recent advancements include Cradle \citep{cradle}, which introduces a General Computer Control setting for broader software interaction, and strategic discussion agents for games like One Night Ultimate Werewolf \citep{jinwerewolf}. Benchmarking platforms like AGENTBENCH \citep{agentbench} play a crucial role in evaluating these developments in comprehensive, open-ended contexts.

\textbf{Human-AI Collaboration in Gaming}: Recent advancements in Large Language Models (LLMs) have opened new avenues for human-AI collaboration in complex gaming environments\citep{feng2024large}. MindAgent \citep{mindagnet} introduces a framework for evaluating AI planning and coordination in multi-agent gaming scenarios, while practical applications like the AI teammate in "Naraka: Bladepoint" mobile game \footnote{\url{https://fuxi.163.com/database/1179}} demonstrate the potential of AI assistants in commercial games. 

These developments showcase progress in AI-assisted gaming and human-AI interaction, yet there remains a gap in leveraging such technologies for real-time strategy assistance in complex RTS games like StarCraft II. Our work aims to address this gap by introducing a language-conditioned, adaptive system that enhances human decision-making in StarCraft II, while also exploring its potential to improve accessibility and overall gaming experience.
\section{Method}
\label{sec:method}

This section presents the architecture and key components of Adaptive Command, our novel system based on TextStarCraft II for enhancing human decision-making in StarCraft II through real-time AI assistance.

\subsection{System Architecture}
\label{subsec:system_architecture}

Adaptive Command consists of five main components:

\begin{enumerate}
    \item \textbf{Game State Processor}: Employs the Chain of Summarization (CoS) method to efficiently process and compress game information.
    \item \textbf{LLM-based Strategic Advisor}: Utilizes pre-trained models like GPT-4 and DeepSeek to analyze game states and generate strategic advice.
    \item \textbf{Behavior Tree Framework}: Translates high-level strategies into executable game actions, controlled by Behavior Modulators (BMs).
    \item \textbf{Natural Language Interface}: Facilitates communication between the player and the AI system through voice commands and responses.
    \item \textbf{Real-time Policy Adjustment Mechanism}: Dynamically updates game strategies by modifying BMs based on player input and game state.
\end{enumerate}

Figure \ref{fig:adaptive_command_framework} illustrates the overall architecture of these components.

\subsection{Mathematical Formulation}
\label{subsec:math_formulation}

We formulate Adaptive Command as a two-phase process within a reinforcement learning framework:

\begin{algorithm}[tb]
    \caption{Adaptive Command in StarCraft II}
    \label{alg:adaptive_command}
    \textbf{Input}: StarCraft II game environment $\mathcal{E}$, language model $f_{LLM}$, behavior tree $g_{BT}$
    \begin{algorithmic}[1]
        \State Initialize game state $s_0 = \mathcal{E}.\text{reset}()$
        \State $c_0 \gets$ initial player instruction
        \State $\pi_0 \gets f_{LLM}(s_0, c_0, \theta)$ \Comment{Initial policy selection}
        \State $t \gets 0$
        \While{$\mathcal{E}$ is not terminated}
            \State $a_t \gets g_{BT}(\pi_t, s_t)$ \Comment{Action selection}
            \State $s_{t+1}, r_t \gets \mathcal{E}.\text{step}(a_t)$ \Comment{Environment interaction}
            \If{player provides new instruction $c_{t+1}$}
                \State $\hat{s}_{t+1} \gets \text{CoS}(s_{t+1})$ \Comment{Chain of Summarization}
                \State $\pi_{t+1} \gets f_{LLM}(\hat{s}_{t+1}, a_t, c_{t+1}, \theta)$ \Comment{Policy adjustment}
            \Else
                \State $\pi_{t+1} \gets \pi_t$
            \EndIf
            \State $t \gets t + 1$
        \EndWhile
    \end{algorithmic}
\end{algorithm}

Where:
\begin{itemize}
    \item $\mathcal{E}$ is the StarCraft II game environment
    \item $s_t \in \mathcal{S}$ is the game state at time step $t$
    \item $a_t \in \mathcal{A}$ is the action taken at time step $t$
    \item $\pi_t$ is the policy at time step $t$
    \item $c_t$ is the player's instruction at time step $t$
    \item $r_t$ is the reward at time step $t$
    \item $\theta$ represents the parameters of the LLM
    \item $f_{LLM}$ is the LLM's decision function
    \item $g_{BT}$ is the behavior tree execution function
    \item $CoS$ is the Chain of Summarization function
\end{itemize}

This formulation captures the two-phase nature of Adaptive Command:
1. Initial Policy Selection (line 3): The LLM selects an initial policy based on the initial game state and player instruction.
2. Real-time Policy Adjustment (lines 8-13): The system adjusts the policy based on new player instructions and the current game state, processed through the Chain of Summarization.

\subsection{Game State Processing}
\label{subsec:game_state_processing}

We adopt the Chain of Summarization (CoS) method, originally introduced in the TextStarCraft II framework \cite{ma2024largelanguagemodelsplay}, for efficient game state processing. This method involves:

\begin{enumerate}
    \item \textbf{Single-Frame Summarization}: Condenses immediate game state information.
    \item \textbf{Multi-Frame Summarization}: Aggregates information over multiple frames to capture temporal dynamics.
    \item \textbf{Strategic Context Integration}: Combines summarized game state with player instructions.
\end{enumerate}

In our algorithm, the CoS function (line 9) processes the raw game state $s_{t+1}$ into a summarized state $\hat{s}_{t+1}$, which serves as input for the LLM. This approach significantly reduces information overload while maintaining strategic clarity, enabling more efficient decision-making processes.

\subsection{LLM Integration}
\label{subsec:llm_integration}

We leverage pre-trained LLMs without additional fine-tuning, allowing them to apply their general language understanding and reasoning capabilities to StarCraft II strategy. The LLM's decision function $f_{LLM}$ is used for both initial policy selection (line 3) and real-time policy adjustment (line 10). It interprets player queries $c_t$, analyzes game states $s_t$ or $\hat{s}_{t+1}$, and generates tactical suggestions in the form of a policy $\pi_t$. The parameters $\theta$ represent the LLM's knowledge and are kept constant during gameplay.

\subsection{Behavior Tree and Modulators}
\label{subsec:behavior_tree}

The Behavior Tree Framework, represented by function $g_{BT}$ (line 6), translates the high-level policy $\pi_t$ into concrete game actions $a_t$. The Behavior Modulators (BMs) serve as the interface between the abstract policy and the specific game commands. These BMs are dynamically adjusted based on the policy $\pi_t$ generated by the LLM and the current game state $s_t$, resulting in actions $a_t$ that are executed in the game environment.

\subsection{Voice Interface}
\label{subsec:voice_interface}

The system incorporates Speech-to-Text (whisper-small) and Text-to-Speech (edge-tts) technologies to enable hands-free interaction. This allows players to input instructions $c_t$ verbally and receive policy suggestions $\pi_t$ as voice output, corresponding to the player input in lines 3 and 8 of our algorithm.

\subsection{StarCraft II Integration}
\label{subsec:sc2_integration}

Adaptive Command integrates with StarCraft II, represented as the environment $\mathcal{E}$ in our algorithm. It receives real-time game state information $s_t$ (line 7) and executes commands $a_t$ through the Behavior Tree Framework, providing a seamless experience without disrupting normal gameplay.

\section{Interaction Pipeline}
\label{sec:interaction_pipeline}

The Adaptive Command system facilitates a dynamic and interactive gameplay experience, combining AI assistance with traditional StarCraft II gameplay. 

\subsection{Initial Policy Selection}
\label{subsec:initial_policy}
At the start of the game (corresponding to lines 1-3 in our algorithm):
\begin{enumerate}
\item The game environment $\mathcal{E}$ is initialized, providing the initial state $s_0$.
\item The player interacts with the LLM through natural language to discuss initial strategy, providing $c_0$.
\item Based on the player's input $c_0$ and initial game state $s_0$, the LLM recommends the most suitable rule-based policy $\pi_0 = f_{LLM}(s_0, c_0, \theta)$.
\item The player can accept the recommendation or request alternatives, refining the strategy through continued dialogue.
\end{enumerate}

\subsection{Real-time Policy Adjustment}
\label{subsec:realtime_adjustment}
During gameplay (corresponding to lines 5-14 in our algorithm):
\begin{enumerate}
\item The system selects actions $a_t = g_{BT}(\pi_t, s_t)$ based on the current policy and game state.
\item The game environment evolves: $s_{t+1}, r_t = \mathcal{E}.\text{step}(a_t)$.
\item If the player provides new instructions $c_{t+1}$:
    \begin{itemize}
        \item The system processes the current game state using the Chain of Summarization: $\hat{s}_{t+1} = \text{CoS}(s_{t+1})$.
        \item The LLM analyzes the summarized game state and player's input to suggest tactical adjustments: $\pi_{t+1} = f_{LLM}(\hat{s}_{t+1}, a_t, c_{t+1}, \theta)$.
    \end{itemize}
\item Upon player approval, the system modifies the behavior tree to implement the new strategy.
\end{enumerate}

\subsection{Voice-based Interaction}
\label{subsec:voice_interaction}

To facilitate hands-free interaction:
\begin{enumerate}
    \item All interactions between the player and the LLM are conducted through a voice interface.
    \item Speech-to-Text (STT) technology converts player's voice commands into text instructions $c_t$ for LLM processing.
    \item Text-to-Speech (TTS) technology converts LLM's policy suggestions $\pi_t$ into voice output for the player.
    \item This enables players to focus on gameplay while discussing strategy, corresponding to the input of $c_t$ and output of $\pi_t$ in our algorithm.
\end{enumerate}

\subsection{Traditional Gameplay Integration}
\label{subsec:traditional_gameplay}
To maintain player agency:
\begin{enumerate}
\item Players retain full control over traditional gameplay mechanics while the AI assistant is available.
\item Keyboard and mouse inputs for direct unit and resource management remain functional, as in standard StarCraft II gameplay.
\item This hybrid approach allows players to leverage both AI strategic assistance $\pi_t$ and their own micro-management skills to influence the final actions $a_t$ executed in the game.
\end{enumerate}

\subsection{Win Conditions}
\label{subsec:win_conditions}
The victory conditions in Adaptive Command remain consistent with standard StarCraft II rules:
\begin{itemize}
\item The primary win condition is the destruction of all enemy buildings.
\item All other standard StarCraft II victory and defeat scenarios apply.
\end{itemize}

This interaction pipeline demonstrates how Adaptive Command seamlessly integrates AI assistance with traditional StarCraft II gameplay, offering players a unique, adaptive, and interactive strategic experience while following the mathematical formulation presented in our algorithm.

\section{Experiment Design and Evaluation}
\begin{figure*}[t]
    \centering
    \begin{subfigure}[b]{0.48\textwidth}
        \centering
        \includegraphics[width=\linewidth]{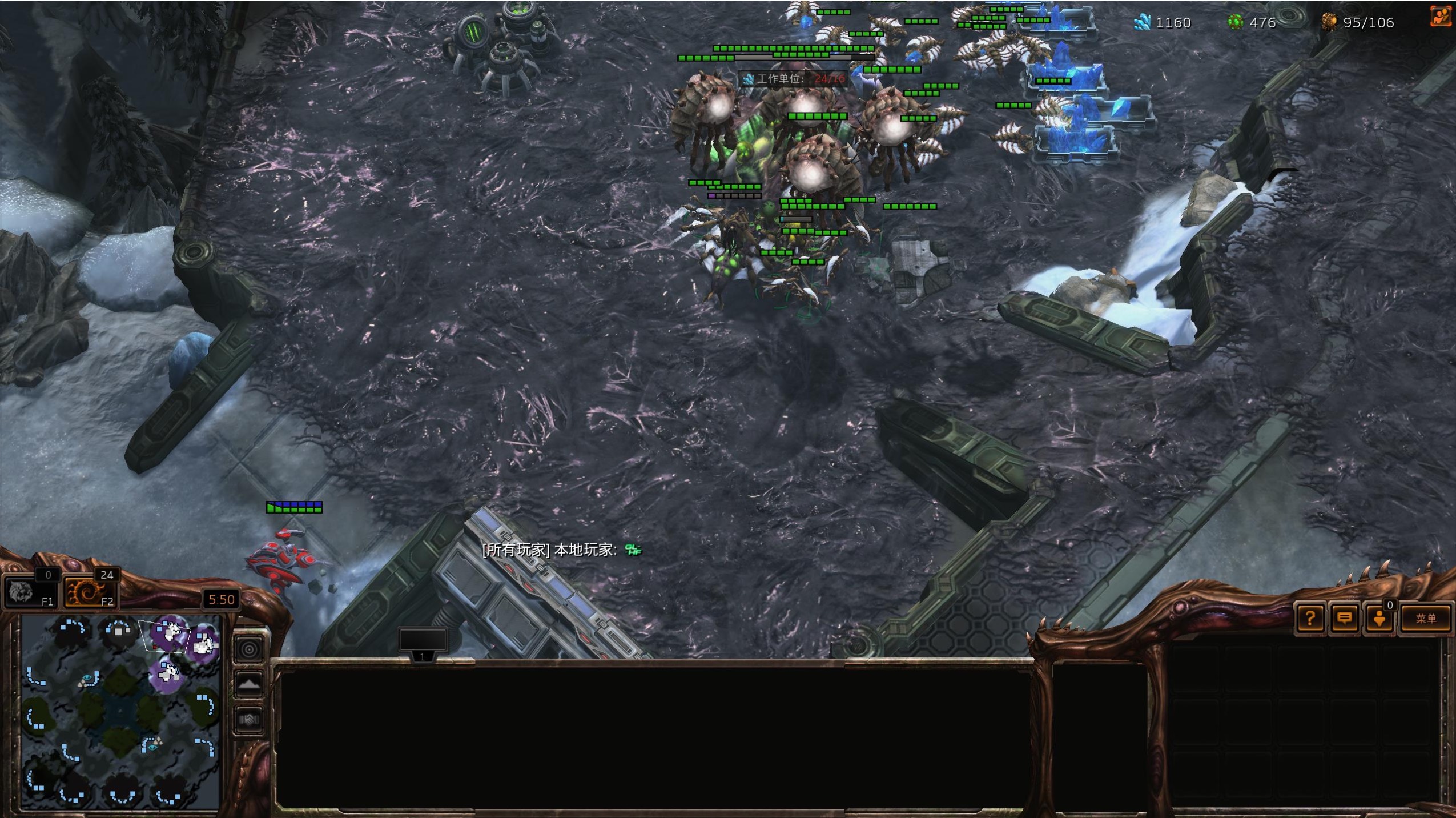}
        \caption{Initial strategy}
        \Description{Screenshot of StarCraft II showing the initial strategy with a mix of ground units.}
        \label{fig:initial}
    \end{subfigure}
    \hfill
    \begin{subfigure}[b]{0.48\textwidth}
        \centering
        \includegraphics[width=\linewidth]{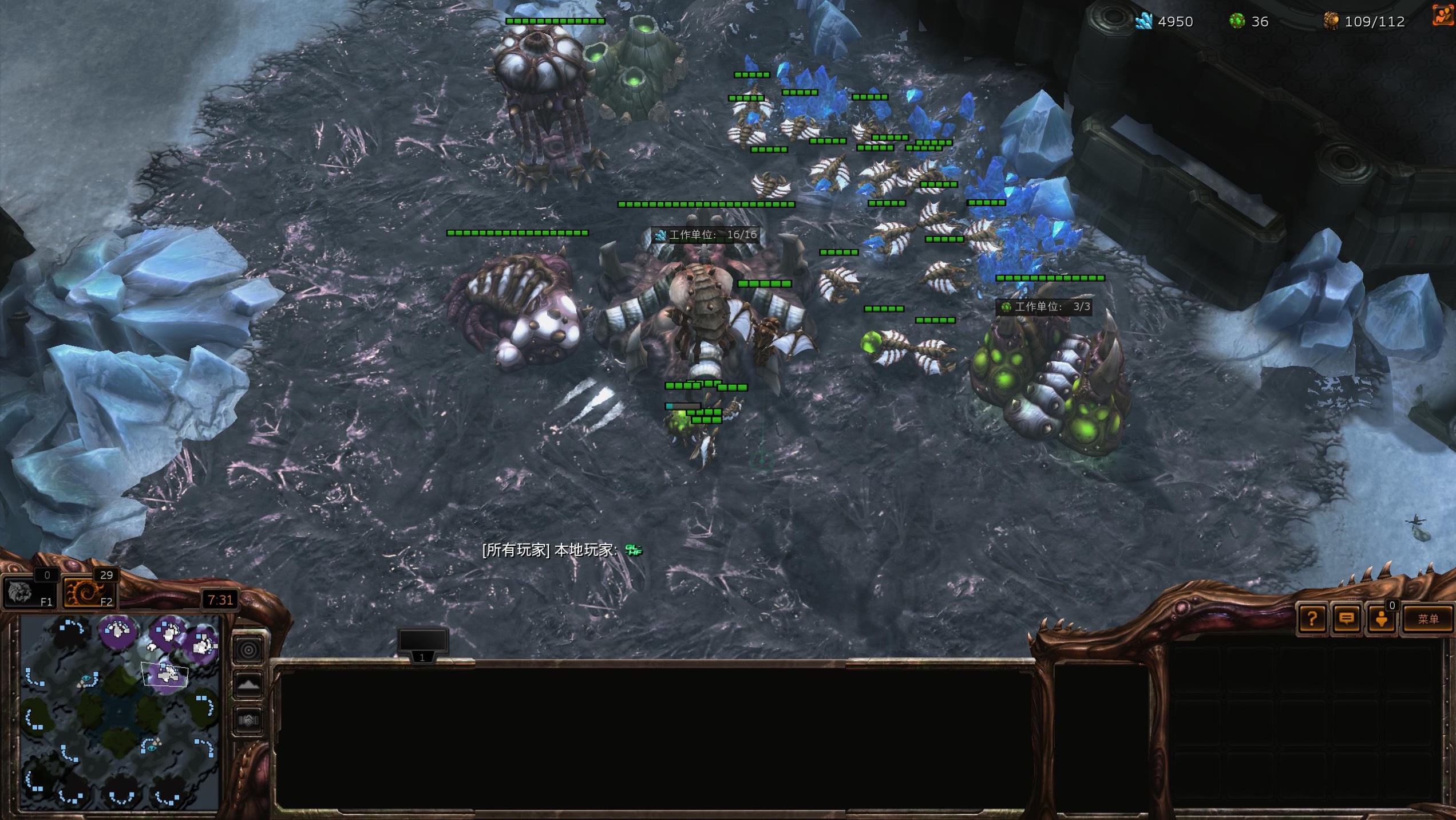}
        \caption{Transition to air units}
        \Description{Screenshot depicting the transition to air units, with Mutalisks and Corruptors visible.}
        \label{fig:air}
    \end{subfigure}
    \vskip\baselineskip
    \begin{subfigure}[b]{0.48\textwidth}
        \centering
        \includegraphics[width=\linewidth]{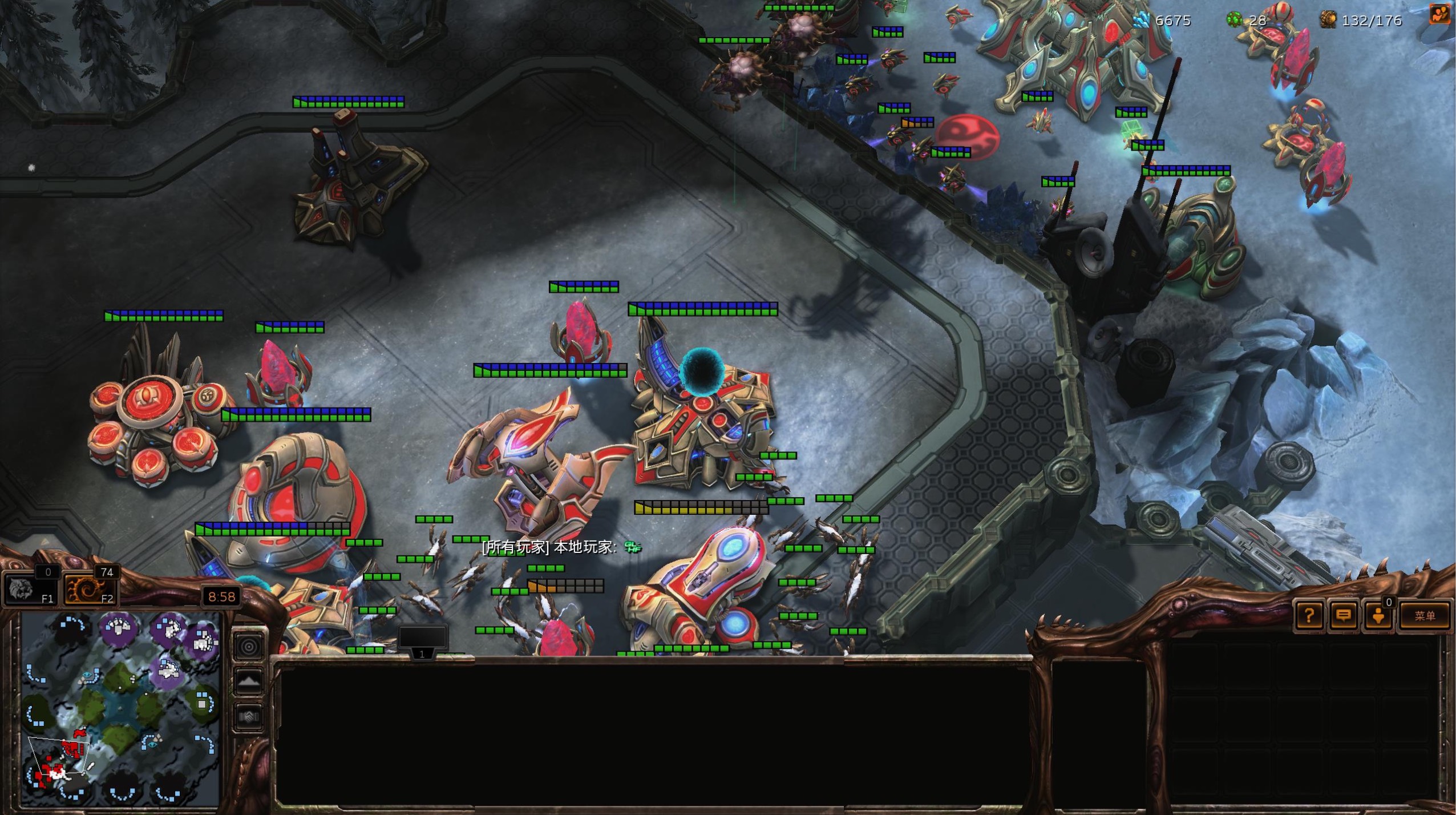}
        \caption{Balanced army composition}
        \Description{Screenshot showing a balanced army composition with both ground and air units.}
        \label{fig:balanced}
    \end{subfigure}
    \hfill
    \begin{subfigure}[b]{0.48\textwidth}
        \centering
        \includegraphics[width=\linewidth]{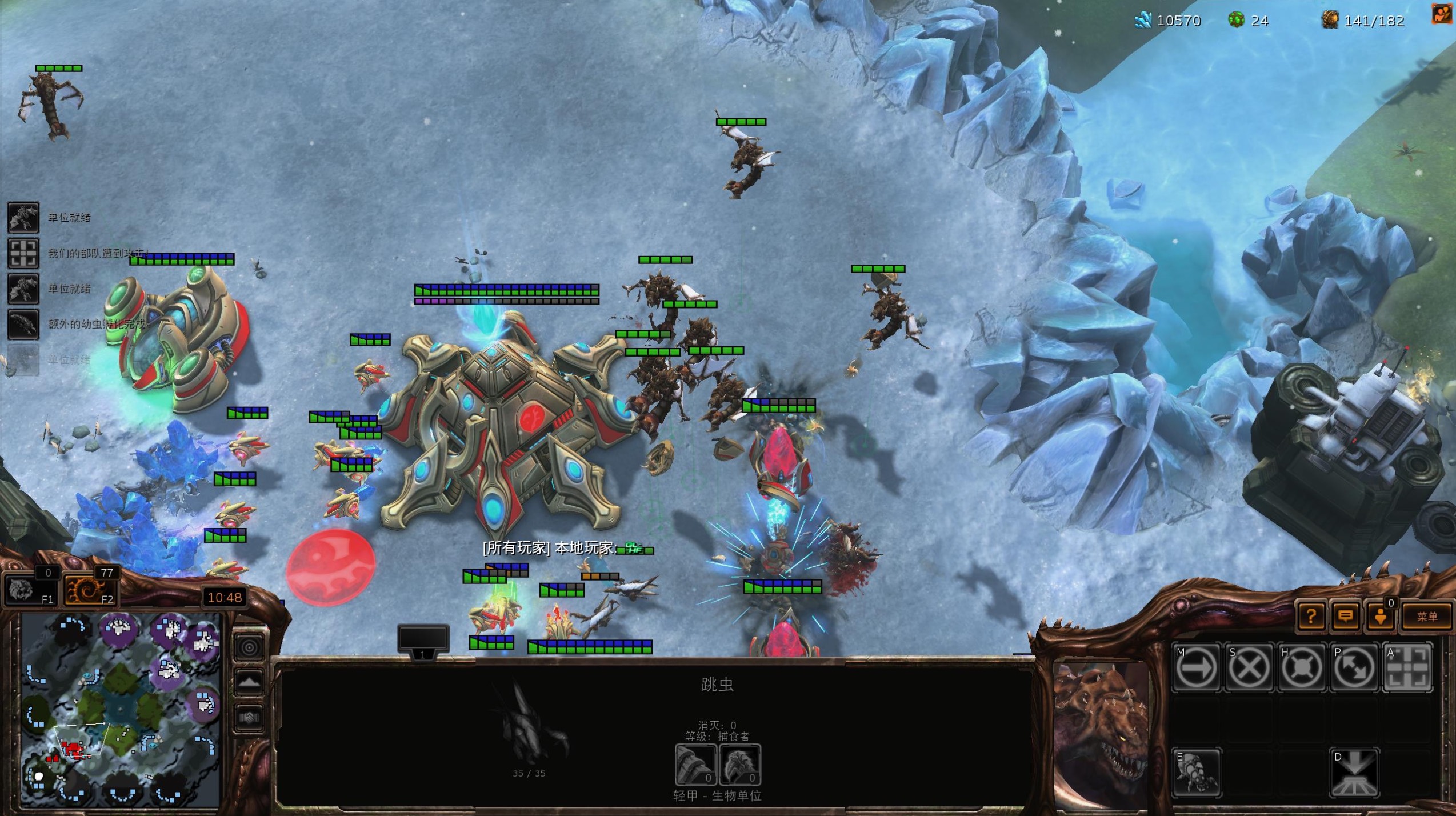}
        \caption{Final victory}
        \Description{Screenshot of the final victory screen, showing the player's triumph.}
        \label{fig:victory}
    \end{subfigure}
    \caption{Adaptive Command in action: Real-time strategy adjustment through human-AI collaboration}
    \label{fig:adaptive_command}
    \Description{A four-part figure showcasing the progression of a StarCraft II game using Adaptive Command. It illustrates the initial strategy, transition to air units, balanced army composition, and final victory. Below the images is a table showing examples of human input and corresponding LLM responses for each phase.}
\begin{tabular}{|p{0.26\textwidth}|p{0.35\textwidth}|p{0.35\textwidth}|}
    \hline
    \textbf{Phase} & \textbf{Human Input} & \textbf{LLM Response} \\
    \hline
    Strategy Init. Phase (Fig. \ref{fig:initial}) & "Given the opponent is using heavily armored units, what is the best counter-strategy?" & Recommends Roach-Hydra composition for balance of survivability and damage against armored units. \\
    \hline
    \multirow{2}{*}{Adaptive Exec. Phase (Fig. \ref{fig:air}, \ref{fig:balanced})} & "I want to play a sky army style" & Suggests transitioning to Mutalisks and Corruptors to counter Protoss air units and control the map. \\
    \cline{2-3}
    & "I think we should also produce some ground army, right?" & Advises maintaining a small ground force for versatility and defense against various Protoss ground units. \\
    \hline
\end{tabular}
\end{figure*}
\subsection{Experimental Setup}

\subsubsection{Participants}
We recruited 12 participants with varying levels of StarCraft II experience through online gaming communities and university networks. Participants were categorized as follows:
\begin{itemize}
    \item 4 novice players (less than 50 hours of gameplay)
    \item 4 intermediate players (50-1000 hours of gameplay)
    \item 4 expert players (over 1000 hours of gameplay and achieved grandmaster level)
\end{itemize}

Table \ref{tab:mmr_table} shows the MMR (Matchmaking Rating) distribution of the participants.

\begin{table}[h!]
  \centering
  \caption{Player MMR in Experiment}
  \label{tab:mmr_table}
  \begin{tabular}{ccc}
    \toprule
    Player ID & Skill Level  & MMR \\
    \midrule
    1-4 & Novice & --- \\
    5 & Intermediate & 1923 \\
    6 & Intermediate & 2599 \\
    7 & Intermediate & 3094 \\
    8 & Intermediate & 3496 \\
    9 & Expert & 5600 \\
    10 & Expert & 5513 \\
    11 & Expert & 5314 \\
    12 & Expert & 5005 \\
    \bottomrule
  \end{tabular}
\end{table}

\subsubsection{Task Design}
Each participant played a total of 6 games as Zerg against the built-in AI (Protoss) opponent set to level 6 (Very Hard). The games were divided into two conditions:
\begin{itemize}
    \item Control condition: 3 games played using standard StarCraft II gameplay
    \item Experimental condition: 3 games played with the assistance of Adaptive Command
\end{itemize}
The order of conditions was counterbalanced across participants to mitigate learning effects.

Games were played on three 2023 ladder maps: 'Altitude LE.SC2Map', 'Gresvan LE.SC2Map', and 'Babylon LE.SC2Map'. These maps were chosen to represent current competitive play environments. There was no time limit imposed on the games.

\subsubsection{System Configuration}
\begin{itemize}
    \item LLM: Gemini 1.5 Flash, DeepSeek-V2-0628, GPT-4-turbo-2024-04-09, GPT-4o-2024-08-06. These models were selected for their high performance and availability as commercial APIs.
    \item Behavior Tree: Custom implementation based on TextStarCraft2, python-sc2\footnote{\url{https://github.com/BurnySc2/python-sc2}}, and the SC2 AI community\footnote{\url{https://aiarena.net/}} resources
\end{itemize}

\subsection{Evaluation Metrics}
We employed both quantitative and qualitative measures to evaluate the effectiveness of Adaptive Command:

\subsubsection{Quantitative Metrics}
\begin{enumerate}
    \item Win Rate: Percentage of games won against the AI opponent in each condition.
    \item Instruction Following: Degree to which the system accurately implemented player instructions in the experimental condition, rated on a 1-5 scale by the research team. A score of 1 indicates complete failure to understand instructions, while 5 indicates perfect comprehension and corresponding adjustments.
\end{enumerate}

\subsubsection{Qualitative Metrics}
\begin{enumerate}
    \item Helpfulness: Players' assessment of how helpful the Adaptive Command system was during gameplay, rated on a 1-5 scale. A score of 1 indicates not helpful at all, while 5 indicates extremely helpful.
\end{enumerate}

\subsection{Limitations}

While our study provides valuable insights into the effectiveness of Adaptive Command, several limitations should be noted:

\begin{itemize}
    \item \textbf{Sample Size:} The study involved only 12 participants, which may limit the generalizability of our findings. A larger sample size could provide more robust and representative results.

    \item \textbf{AI Opponent:} Our experiments were conducted against the built-in AI of StarCraft II. While this ensures consistency across trials, it may not fully capture the complexity and unpredictability of human opponents in real competitive scenarios.

    \item \textbf{Game Version Discrepancy:} There exists a potential mismatch between the game version used during the pre-training of the large language models and the version used in our current experiments. This discrepancy may affect the relevance and accuracy of the AI's strategic advice, as the game's meta and balance may have evolved.

    \item \textbf{Framework Performance Limitations:} The Adaptive Command system relies on a behavior tree for execution, which may inherently limit its performance compared to more advanced reinforcement learning (RL) approaches. While our framework offers interpretability and ease of modification, it may not achieve the same level of optimization and adaptability as RL-based systems.

\end{itemize}

\section{Result Analysis}

\subsection{Quantitative Results}

\subsubsection{Win Rates}
\begin{figure}[t]
    \centering
    \includegraphics[width=\linewidth]{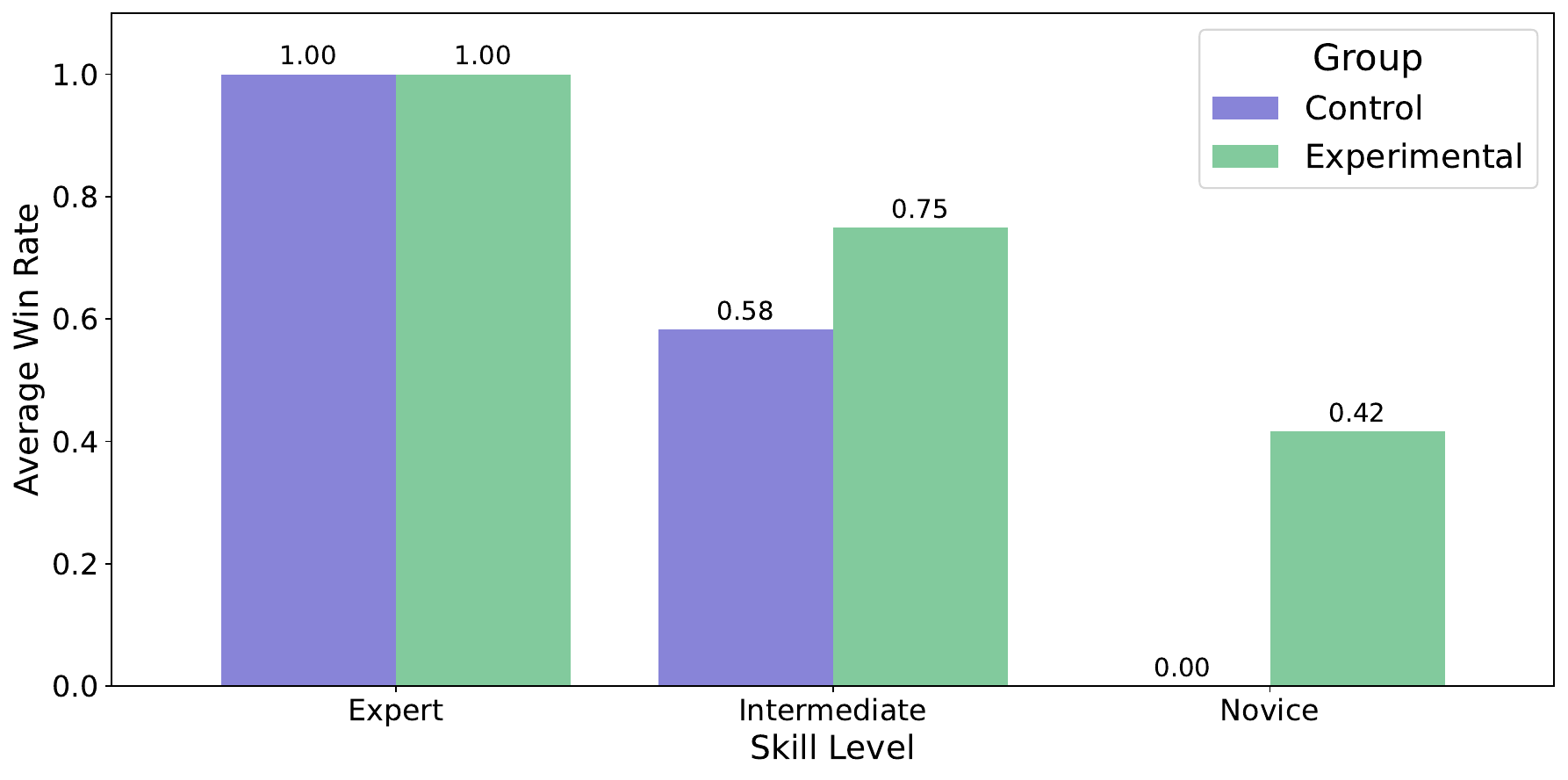} 
    \caption{Average Win Rate by Skill Level and Group} 
    \label{fig:skilllevelperformance}
    \Description{Bar chart comparing the average win rates of players at different skill levels (Novice, Intermediate, Expert) in two groups: Control (standard gameplay) and Experimental (using Adaptive Command). The chart shows significant improvements in win rates for Novice and Intermediate players when using Adaptive Command, while Expert players maintain high win rates in both conditions.}
\end{figure}

We observed significant differences in win rates between the control group (standard gameplay) and the experimental group (using Adaptive Command). The percentages represent the proportion of games won against the level 6 (Very Hard) AI opponent:

\begin{itemize}
    \item Novice players: 0\% (control) vs. 42\% (experimental)
    \item Intermediate players: 58\% (control) vs. 75\% (experimental)
    \item Expert players: 100\% (control) vs. 100\% (experimental)
\end{itemize}

These results suggest that Adaptive Command provides the most substantial benefit to novice players, with a 42 percentage point increase in win rate. The impact on intermediate players is also notable, with a 17 percentage point improvement, while expert players maintained their perfect performance in both conditions.

\subsubsection{Instruction Following and Helpfulness}
\begin{figure}[htbp]
    \centering
    \includegraphics[width=\linewidth]{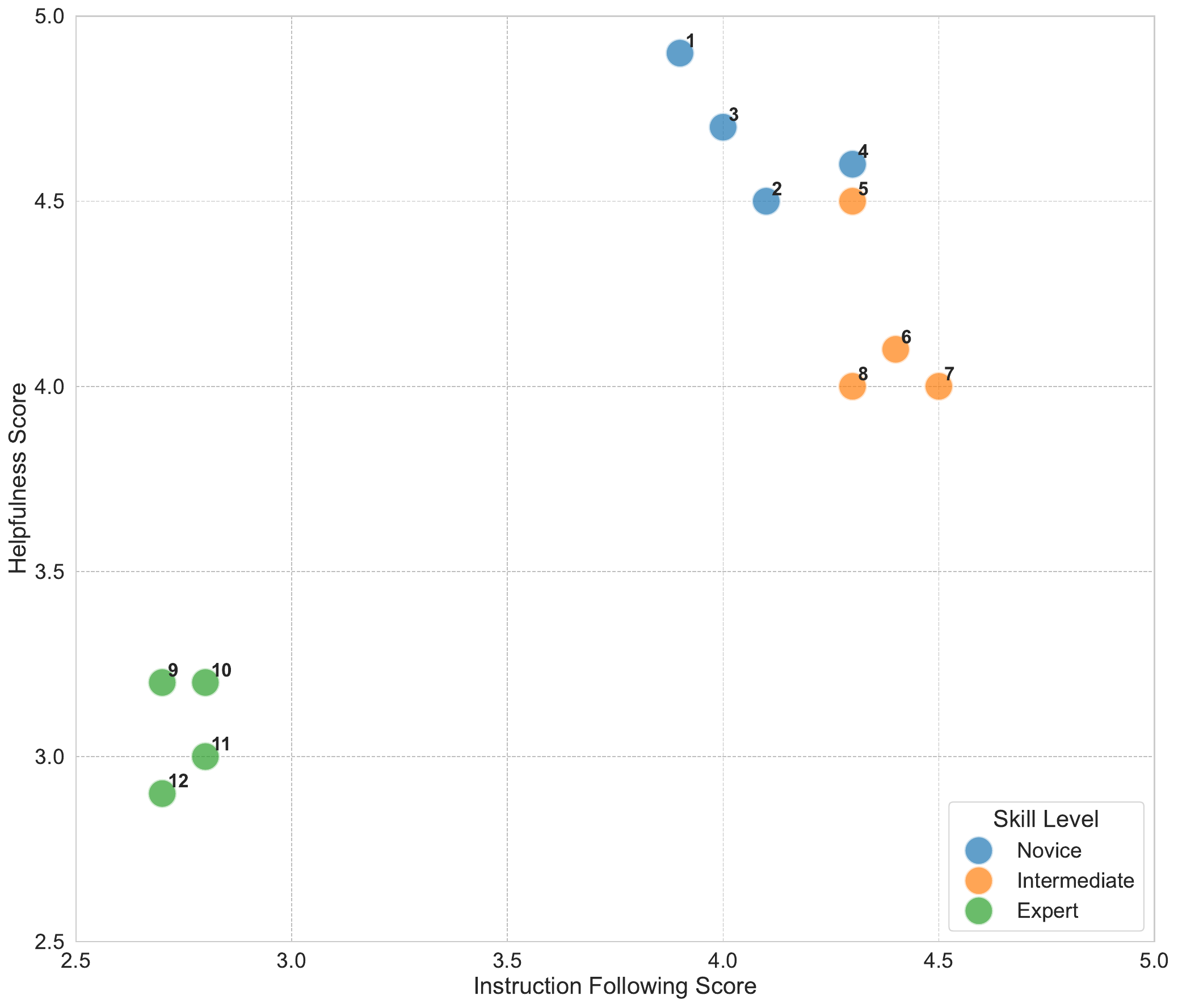}
    \caption{Individual participant scores for Instruction Following and Helpfulness. Each point represents a participant, with numbers indicating participant IDs. Point size and color denote skill level (Novice, Intermediate, Expert). Instruction Following measures the system's ability to accurately implement player instructions, while Helpfulness reflects the perceived usefulness of the Adaptive Command system.}
    \label{fig:individual_scores}
    \Description{test}
\end{figure}

The average instruction following and helpfulness ratings (on a 1-5 scale) for Adaptive Command were:

\begin{itemize}
    \item Novice players: Instruction Following 4.1, Helpfulness 4.7
    \item Intermediate players: Instruction Following 4.4, Helpfulness 4.2
    \item Expert players: Instruction Following 2.8, Helpfulness 3.1
\end{itemize}

Novice and intermediate players reported high scores for both metrics, indicating that the system accurately implemented their instructions and was perceived as highly helpful. Interestingly, expert players reported significantly lower scores for both metrics, suggesting that the system may not have met their more advanced needs or expectations.

\subsection{Discussion of Results}

The results demonstrate that Adaptive Command provides varying benefits across different skill levels:

\begin{itemize}
    \item For novice players, the system serves as a crucial learning tool, providing basic automated strategies that significantly improve their performance. The dramatic increase in win rate (from 0\% to 42\%) and high helpfulness score (4.7) reflect the system's effectiveness in bridging the knowledge gap for beginners.
    
    \item Intermediate players also benefit substantially, with improved win rates (from 58\% to 75\%) and high instruction following (4.4) and helpfulness (4.2) scores. This suggests that the system effectively helps translate their theoretical knowledge into practical gameplay strategies.
    
    \item Expert players, while maintaining perfect win rates, reported lower scores in instruction following (2.8) and helpfulness (3.1). This indicates challenges in meeting the complex needs of high-level play, possibly due to the sophisticated nature of expert strategies and instructions.
\end{itemize}

These findings highlight the system's potential as a skill development tool, particularly for novice and intermediate players, while also revealing areas for improvement in supporting expert-level gameplay.

\section{Discussion, Future Work, and Conclusion}

\subsection{Discussion}

Our study on Adaptive Command reveals significant insights into the potential of Large Language Models (LLMs) in enhancing human-AI collaboration in complex gaming environments:

\begin{itemize}
    \item \textbf{Skill Development Tool}: The system proves highly effective for novice and intermediate players, serving as both a learning platform and a performance enhancer. This demonstrates the potential of AI assistants in accelerating skill acquisition in complex games.
    
    \item \textbf{Cognitive Load Reduction}: Across all skill levels, Adaptive Command helps reduce the cognitive load associated with macro management. This allows players to focus more on tactical decision-making and micro-management, potentially leading to more engaging and strategic gameplay experiences.
    
    \item \textbf{Language-Conditioned Policy Effectiveness}: The high instruction following ratings, particularly for novice and intermediate players, showcase the system's ability to translate natural language instructions into game actions. This opens up new possibilities for intuitive human-AI interaction in gaming.
    
    \item \textbf{Scalability Challenges}: The disparity in effectiveness across skill levels, particularly the lower scores from expert players, highlights the challenges in creating AI systems that can cater to all levels of expertise. This points to the need for more advanced, flexible AI models that can adapt to varying levels of player sophistication.
\end{itemize}

\subsection{Future Work}

Building upon these insights, our future research will focus on:

\begin{itemize}
    \item Integrating LLMs with reinforcement learning models to create a more adaptive and high-performance system.
    \item Developing more sophisticated natural language processing capabilities to better interpret and execute complex, expert-level instructions.
    \item Exploring ways to dynamically adjust the level of AI assistance based on real-time assessment of player skill and needs.
    \item Investigating the application of this technology to other complex, decision-intensive environments beyond gaming.
\end{itemize}

We anticipate that these advancements will not only enhance gaming experiences but also contribute to broader applications of human-AI collaboration in fields such as education, business strategy, and defense.

\subsection{Conclusion}

Adaptive Command represents a significant step forward in creating language-conditioned AI assistants for complex decision-making environments. While current limitations exist, particularly in matching expert-level play, the system demonstrates remarkable potential in:

\begin{itemize}
    \item Accelerating skill development for novice and intermediate players.
    \item Providing a novel approach to human-AI collaboration in real-time strategy games.
    \item Offering insights into how language models can enhance human decision-making in complex, time-sensitive scenarios.
\end{itemize}

This research not only contributes to the field of game AI but also opens up new avenues for exploring human-AI collaboration in various industries. As we continue to refine and expand this technology, we anticipate significant advancements in our understanding and application of AI-assisted decision-making in complex, real-world environments.

\section{Appendix: Introduction to StarCraft II}
\label{appendix:introduction of starcraft2}

\textbf{StarCraft II (SC2)} is a real-time strategy (RTS) game developed by Blizzard Entertainment, renowned for its strategic depth and complexity. It serves as both a popular e-sport and a benchmark for AI research in strategic gameplay.

\subsection{Gameplay Overview}
Players choose from three distinct species: Terrans, Protoss, and Zerg, each offering unique units and strategies. The game revolves around four key elements:

\begin{enumerate}
    \item \textbf{Resource Management}: Gathering minerals and vespene gas to fund operations.
    \item \textbf{Base Construction}: Expanding and fortifying positions efficiently.
    \item \textbf{Army Composition}: Building and controlling a balanced mix of units.
    \item \textbf{Strategic Planning}: Adapting strategies in response to opponent moves.
\end{enumerate}

\section{ Introduction to StarCraft II}
\label{appendix:introduction of starcraft2}

\textbf{StarCraft II (SC2)} is a real-time strategy (RTS) game developed by Blizzard Entertainment, renowned for its strategic depth and complexity. It serves as both a popular e-sport and a benchmark for AI research in strategic gameplay.

\subsection{Gameplay Overview}
Players choose from three distinct species: Terrans, Protoss, and Zerg, each offering unique units and strategies. The game revolves around four key elements:

\begin{enumerate}
    \item \textbf{Resource Management}: Gathering minerals and vespene gas to fund operations.
    \item \textbf{Base Construction}: Expanding and fortifying positions efficiently.
    \item \textbf{Army Composition}: Building and controlling a balanced mix of units.
    \item \textbf{Strategic Planning}: Adapting strategies in response to opponent moves.
\end{enumerate}

\subsection{Competitive Scene and Research Platform}
Since its 2010 launch, SC2 has been a cornerstone of e-sports, featuring in major tournaments worldwide. Its demanding mechanics and strategic complexity have also made it a valuable platform for AI research, particularly in areas of real-time decision making, strategic planning, and human-AI collaboration.

The game's rich strategic landscape, coupled with its clear victory conditions and quantifiable performance metrics, makes it an ideal testbed for developing and evaluating AI systems capable of complex, multi-step reasoning and execution in dynamic environments.
\section{Acknowledgements}
This work was supported by the National Science and Technology Major Project 2022ZD0116404.

\bibliographystyle{ACM-Reference-Format}
\bibliography{sample-base}

\end{document}